\documentclass[12pt]{article}

\usepackage{graphicx}

\begin{document}
\def\Dirac{i\partial\!\!\!\!/}
\def\DDirac{iD\!\!\!\!/}
\def\dirac{i\partial\!\!\!\!/-eA\!\!\!\!/}
\newcommand{\beq}{\begin{eqnarray}}
\newcommand{\eeq}{\end{eqnarray}}
\newcommand{\nn}{\nonumber}
\newcommand{\rparen}{({\bf r})}
\newcommand{\kparen}{({\bf k})}
\newcommand{\sumn}{\sum\limits}
\newcommand{\bfr}{{\bf r}}
\newcommand{\bfk}{{\bf k}}
\newcommand{\bfK}{{\bf K}}
\newcommand{\bfp}{{\bf p}}
\newcommand{\bfKpm}{{\bf K_{\pm}}}
\newcommand{\varpm}{{\varphi_{\pm}}}
\newcommand{\apm}{{a_{\pm}}}
\newcommand{\bpm}{{b_{\pm}}}

\begin{center}
{\Large\bf \boldmath Phases of the Dirac determinant, Abelian Chern-Simons terms and Berry's phases in the field theoretic description of graphene
} 

\vspace*{6mm}
{Eve Mariel Santangelo$^{a,b}$ }\\      
{\small \it $^a$Departamento de F\'{\i}sica, Universidad Nacional de La Plata, Argentina \\
            $^b$Instituto de F\'{\i}sica de La Plata, CONICET and Universidad Nacional de La Plata, Argentina}
\end{center}

\vspace*{6mm}

\begin{abstract}

Abstract: This talk presents a study of massless relativistic Dirac fields in three Euclidean dimensions, at finite temperature and density, in the presence of a uniform electromagnetic background. Apart from explaining the behavior of Hall's conductivity for graphene, our results show a direct relationship between the selection of a phase for the Dirac determinant and the generation (or lack thereof) of Berry's phases and Chern-Simons terms.

\end{abstract}

\vspace*{6mm}

\section{Introduction}\label{sect1}

Graphene is a bidimensional array of carbon atoms, packed in a
honeycomb crystal structure. Even though its theoretical properties were studied decades ago \cite{mele}, it was only in 2005
that stable monolayer samples of such material were obtained. Among other properties, the Hall conductivity was measured in such samples, independently, by two groups \cite{nature1}. Later on, a different behavior of the Hall conductivity was reported \cite{novo2} for bilayer
samples. The main difference between the behavior of the Hall conductivity of mono- and bilayer samples lies in the height of the jump around zero carrier density (or, equivalently, around zero chemical potential).

From a theoretical point of view, the most remarkable feature of graphene is that, in a small momentum approximation, the charge carriers or {\sl
quasi}--particles behave as two ``flavors" (to account for the
spin of the elementary constituents) of massless
relativistic Dirac particles in the two non--equivalent
representations of the Clifford algebra (corresponding to the two
non--equivalent vertices in the first Brillouin zone), with an effective ``speed of light"
about two orders of magnitude smaller than $c$ \cite{mele}.

In \cite{BS1}, we showed that a field theoretic
calculation at finite temperature and density, based upon $\zeta-$function regularization of the
Dirac determinant leads, in the zero temperature limit, to a sequence of plateaux in the Hall conductivity consistent with the ones measured each time the chemical potential goes through a nonzero Landau level. Moreover, it was shown in \cite{bgss} that two of the three possible combinations of phases of the Dirac determinant in both nonequivalent Clifford representations predict a behavior around zero chemical potential consistent with the ones measured in mono- and bilayer graphene.

This paper presents, in section \ref{sect3}, a brief review of our previous results on the subject, with emphasis on the role of the phase of the determinant in giving rise to different behaviors of the Hall conductivity around zero chemical potential. In section \ref{sect4}, we allow for complex chemical potentials, and concentrate on the contribution due to the lowest Landau level, in order to study the invariance of the effective action under large gauge transformations, i.e., under statistics-preserving transformations. We also discuss the connection among phases of the determinant, Berry's phases and Chern-Simons terms.

\section{The Hall conductivity and its dependence on the phase of the determinant}\label{sect3}

As shown in our previous work on the subject \cite{BS1,bgss}, the Hall conductivity can be determined by first evaluating the partition function (equivalently, the effective action) for massless Dirac fermions at finite temperature and density, in two spacial dimensions, in the presence of an external magnetic field perpendicular to the plane, and then performing a boost to a reference frame with orthogonal electric and magnetic fields. In this section, we merely list our main results in those references, with emphasis on the role played by the phase of the Dirac determinant, which appears when treating the infinite tower of states associated to the lowest Landau level. We first consider a single flavor, and one of the two nonequivalent representations of the Clifford algebra.

In order to take into account the effects due to finite temperature and density, we study the theory in Euclidean three-dimensional
space, with a compact Euclidean ``time" $0\leq x_0\leq \beta$, where $\beta=\frac{1}{k_BT}$ (here, $k_B$ is the Boltzmann constant and $T$ is the temperature). We introduce the (real) chemical potential and the magnetic field through a minimal coupling of the theory to an electromagnetic potential  $A_{\mu}=(-i\frac{\mu}{e},0, Bx_1)$. Natural units ($c=\hbar=1$) will be used, unless otherwise stated.

In this scenario, the Euclidean effective action is given by $\log{\cal Z}=\log\, det(\dirac)_{AP}$, where the subindex $AP$ indicates that antiperiodic boundary conditions must be imposed in the $x_0$ direction, in order to ensure Fermi statistics. Now, this is a formal expression, which we will define through a zeta-function regularization, i.e.,
\beq\left.S_{eff}=\log{{\cal Z}}\equiv-\frac{d}{ds}\right\rfloor_{s=0}\,\zeta(s,\frac{(\dirac)_{AP}}
{\alpha})=\left.-\frac{d}{ds}\right\rfloor_{s=0}\sum_{\omega}{\left(\frac{\omega}{\alpha}\right)}^{-s}\,,\label{zeta}\eeq
where $\omega$ represents the eigenvalues of the Dirac operator acting on antiperiodic, square-integrable functions, and $\alpha$ is a parameter introduced to render the zeta function dimensionless (as expected on physical grounds, our final predictions will be $\alpha$-independent).

So, in order order to evaluate the partition function, we first determine the eigenfunctions,
 and the corresponding eigenvalues, of the Dirac operator. We propose
 \[\Psi_{k,l}({x_0},x_1,x_2)=\frac{e^{i\lambda_l {x_0}}e^{ikx_2}}{\sqrt{2\pi\beta}}\,\left(\begin{array}{c} \varphi_{k,l} (x_1) \\\chi_{k,l} (x_1) \\\end{array}\right) \quad\lambda_l=(2l+1)\frac{\pi}{\beta}\,.\]

Note that, in the last expression, $\lambda_l, \, l=-\infty,...,\infty$ are the Matsubara frequencies adequate to the required antiperiodic conditions, while the continuous index $k$ represents an infinite degeneracy in the $x_2$ direction.

The resulting spectrum has two pieces:
An asymmetric piece, associated to the lowest Landau level of the Hamiltonian:
\[
\omega_l=\tilde\lambda_l,  \quad{\rm with}\quad \tilde\lambda_l=(2l+1)\frac{\pi}{\beta}+i\mu \quad {\rm and}\quad l=-\infty,...,\infty\,,\]
and a symmetric piece
\[
\omega_{l,n}=\pm \sqrt{{\tilde \lambda_l}^2+2neB}\quad {\rm with}\quad n=1,...,\infty \quad  l=-\infty,...,\infty\,,\]
corresponding to eigenfunctions with both components different from zero. In all cases, the degeneracy per unit area is given by the well known Landau factor, $\Delta_L\ =\ \frac{eB}{2\pi}$.

The asymmetric part of the spectrum is quite particular. In fact, the corresponding eigenfunction is an eigenfunction of the Pauli matrix $\sigma_3$, with eigenvalue $+1$. The eigenfunction with the opposite ``chirality" was eliminated by the square integrability condition in $x_1$. As we will discuss in what follows, this part of the spectrum is the one which requires the consideration of a phase of the determinant when evaluating the effective action. Before going to such evaluation, it is interesting to note the invariance of the whole spectrum under $\mu\rightarrow \mu+\frac{2ik\pi}{\beta}$. This invariance is a natural one, since such transformations preserve the antiperiodicity of the eigenfunctions and, thus, the Dirac statistics. They are nothing but the so-called large gauge transformations. We will discuss this point in more detail in section \ref{sect4}.

As is clear from (\ref{zeta}), in evaluating the effective action, one must perform the analytic extension of the contributions to the zeta function coming from the nonsymmetric piece of the spectrum, ${\zeta}_1 (s,\mu)$, and the one due to the symmetric piece, ${\zeta}_2(s,\mu,eB)$.

The analytic extension of ${\zeta}_2(s,\mu,eB)$ is quite standard, and it relies mainly on performing a Mellin transform and making use of the inversion properties of the Jacobi theta functions. A detailed presentation can be found in \cite{BS1}.

As said before, the extension of ${\zeta}_1(s,\mu,eB)$ requires a careful consideration of the phase of the determinant. In fact, ${\zeta}_1$ can be written as
\beq{\zeta}_1 (s,\mu)=\Delta_L \left(\frac{2\pi}{\alpha\beta}\right)^{-s}\left[ \sum_{l=0 }^{\infty}\left[(l+\frac12)+i\frac{\mu\beta}{2\pi}\right]^{-s}+\sum_{l=0 }^{\infty}\left[-\left((l+\frac12)-i\frac{\mu\beta}{2\pi}\right)\right]^{-s}\right]\,,\label{splitting}\eeq
and the definition of the overall minus sign in the second sum depends on the selection of the cut in the complex plane of eigenvalues. As discussed in detail in \cite{bgss}, the usual prescription is to choose the cut such that one does not go through vanishing arguments when continuously transforming eigenvalues with positive real part into eigenvalues with negative real part \cite{ecz}. This prescription then gives rise to what will be called in the following the standard phase of the determinant (characterized from now on by $\kappa=-1$). One could certainly choose the opposite prescription, which we will call the nonstandard phase ($\kappa=+1$).
Once one of the phases is selected, the contribution of  ${\zeta}_1$ to the effective action can be evaluated by making use of the well-known properties of the Hurwitz zeta function, to obtain
\[S_{eff}^{I}(\kappa)=\Delta_L
 \left\{\log{\left[2\cosh(\frac{\mu\beta}{2})\right]}+ \kappa\frac{|\mu| \beta}{2}\right\}\,.\]

When this last contribution is added to the one coming from ${\zeta}_2(s,\mu,eB)$, one gets for the effective action
\[S_{eff}(\kappa)=\!\Delta_L\!
 \left\{\log{\left[2\cosh(\frac{\mu\beta}{2})\right]}+ \kappa\frac{|\mu| \beta}{2}+\beta\sqrt{2e B}\zeta_R \left(-\frac{1}{2}\right)\right.\\\]
\[+ \left.\sum_{n=1}^{\infty} \log{\left[\left(1+
e^{-(\sqrt{2ne B}-\mu)\beta}\right)\left(1+
e^{-(\sqrt{2ne B}+\mu)\beta}\right)\right]}\right\}\,.\]

From this last expression, the finite-temperature charge density can be obtained as $ j^0(\kappa)=\frac{-e}{\beta}\frac{d}{d\mu}S_{eff}(\kappa)$. In the zero-temperature limit ($\beta\rightarrow\infty$), and recovering physical units, it reduces to
\[ j^0(2e c^2\hbar Bn<{\mu}^2 < 2e B
c^2\hbar(n+1))=\frac{-(n+\frac{1+\kappa}{2})ce^2B}{h}\, sign(\mu)\,,\]
where $n=[\frac{\mu^2}{2eB\hbar c}]$, and $[x]$ is the integer part of $x$.

In order to obtain the Hall conductivity, one must perform a boost to a reference frame with crossed electric and magnetic fields. The final contribution to the Hall conductivity from each fermion species and one irreducible representation is given by \cite{bgss}
$\sigma_{xy}=\frac{-(n+\frac{1+\kappa}{2})e^2}{h}\, sign(\mu)\,.$

\begin{figure}
\centerline{\includegraphics{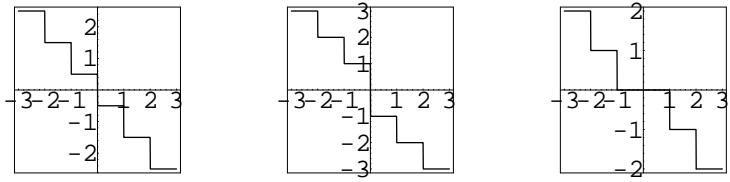}}
\caption{Hall conductivity for different selections of the phase of the determinant.
Left to right: $K=1\,,$ $K=2\,,$ $K=0\,.$ In all cases, the horizontal axis represents $\nu_C={\rm sgn}(\mu)\,\mu^2/{2eB\hbar c^2}\,$
and the vertical one, ${\sigma_{xy}\,h}/{4e^2}\,.$}
\label{figura2}
\end{figure}

Now, the phases of the determinant in both irreducible representations can be selected with the same or with opposite criteria. When this is taken into account, and an overall factor of $2$ is included, to take both fermion species into account, on obtains for the total zero-temperature Hall conductivity
\[\sigma_{xy}=\frac{-4(n+\frac{K}{2})e^2}{h}\, sign(\mu)\,,\]
where $K=0$ corresponds to selecting the standard phase of the determinant in both irreducible representations, $K=1$ corresponds to choosing opposite criteria for the phases, and $K=2$, to choosing both phases in the nonstandard way. The dependence of the Hall conductivity on the classical filling factor ($\nu_C$) is presented in figure \ref{figura2}, for the three values of $K$. From that figure, it is clear that the behavior of monolayer graphene, as reported in \cite{nature1}, corresponds to $K=1$, i.e., to choosing opposite phases of the determinant in both representations. In fact, in this case the (rescaled) Hall conductivity shows a jump of height $1$ for $\nu_C=0$, and further jumps of the same magnitude for $\nu_C=\pm1,\pm2,...$. In turn, the behavior of bilayer graphene, as reported in \cite{novo2} is exactly reproduced by $K=2$ (nonstandard selection of the phase in both representations).

\section{Phases of the determinant as geometric phases}
\label{sect4}

To analyze the physical meaning of the invariance of the effective action under large gauge transformations in this context,
we go back to the zeta function associated to the asymmetric part of the spectrum, for one fermion species and one representation, this time allowing for an imaginary part in the chemical potential, $\tilde{\mu}=\mu+i\,\gamma$, while always keeping $\mu\neq 0$.
In this case, one must be careful when splitting the infinite sum as in (\ref{splitting}). In fact, such splitting must be different for different $\gamma$-ranges, to make sure that all the eigenvalues in each infinite sum have a real part with the same sign, which is crucial in defining the phase. For example, for $-\frac12 <\frac{\gamma \beta}{2\pi} <\frac12$, one has
\beq
 S_{eff}^{I}(-\frac12 <\frac{\gamma \beta}{2\pi} <\frac12)&=&
\left. -\Delta_L\,\frac{d}{ds}\right\rfloor_{s=0}
\left\{\sum_{l=0}^{\infty}\left[\,(2l+1)
{\pi}/{\beta}+i\mu-\gamma\,\right]^{-s}\right.\nn\\
&+&\left.\sum_{l=0}^{\infty}e^{-is\,\theta}\left[\,(2l+1){\pi}/{\beta}
+i(\mu+i\gamma)\,e^{-i\theta}\,\right]^{-s}\right\}\ .\nn
\label{complex}
\eeq

Now, the values of $\theta$ such that the second term in the RHS
does vanish are those ones for which, simultaneously, $(2l+1){\pi}/{\beta}+\mu \sin{\theta}-\gamma \cos{\theta}\ =\ 0\ =
\mu \cos{\theta}+\gamma \sin{\theta}$.

As before, we consider here two different definitions of the phase of the determinant, which correspond to the standard
definition for the phase $\kappa=-1\,$, and to the nonstandard one
$\kappa=+1$. With each one of these prescriptions,
the contribution of the asymmetric spectrum to the effective a action in this range is given by
\beq
S_{eff}^{I}(-\frac12 <\frac{\gamma \beta}{2\pi} <\frac12)
=\Delta_L\left\{\frac{(\kappa+1)\beta}{2}\,{\rm sgn}\,\mu\,(\mu+i\gamma)\right.\nn \\ \!\!\!+\left.\log\left(e^{-\frac{\beta}{2}(\mu+i\gamma)(1+{\rm sgn}\,\mu)}+e^{\frac{\beta}{2}(\mu+i\gamma)(1-{\rm sgn}\,\mu)}\right)
\right\}.\label{mpipi}\eeq

Things are entirely different for $\frac{\gamma \beta}{2\pi} = \pm \frac12$. In this case, one mode in the infinite sum defining the zeta function has a vanishing real part. A careful treatment shows that, at such points, $S_{eff}^{I}$ is discontinuous. For instance, $S_{eff}^{I}(\frac{\gamma \beta}{2\pi} = + \frac12)$ coincides with $\lim_{\frac{\gamma \beta}{2\pi}\rightarrow {\frac12}^{-}}$ of (\ref{mpipi}). An equally carefully treatment of the case $\frac{\gamma \beta}{2\pi}=-\frac12$ shows that $S_{eff}^{I}(\frac{\gamma \beta}{2\pi}=-\frac12)=S_{eff}^{I}(\frac{\gamma \beta}{2\pi}=\frac12)$. This analysis can be extended to other ranges of variation of $\frac{\gamma \beta}{2\pi}$, to obtain
\beq  S_{eff}^{I}((k-\frac12)<\frac{\gamma \beta}{2\pi} &\leq& (k+\frac12))=
\Delta_L\left\{\frac{(\kappa+1)\beta}{2}\,{\rm sgn}\,
\mu\,[\mu+i(\gamma-\frac{2k\pi}{\beta})]\right.\nn \\
&+&\left.\log\left(e^{-\frac{\beta}{2}(\mu+i(\gamma-\frac{2k\pi}{\beta}))(1+{\rm
sgn}\,\mu)}+e^{\frac{\beta}{2}(\mu+i(\gamma-\frac{2k\pi}{\beta}))(1-{\rm
sgn}\,\mu)}\right) \right\},\label{general}\eeq
for $k=-\infty,...,\infty$.

This expression shows that the contribution to the effective
action of the nonsymmetric part of the spectrum, in this
representation of the gamma matrices, is invariant under large
gauge transformations, no matter which phase of the determinant is
selected. As already said, such transformations must constitute an invariance. In fact, an increase of $i\gamma$ in the chemical potential corresponds to the multiplication of the eigenfunctions with a phase, i.e., $\psi_{k,\,l}(x)\rightarrow e^{i\gamma x_0}\psi_{k,\,l}(x)$. So, an increase $i\gamma=\frac{2i\pi}{\beta}$ is a pure gauge transformation which, moreover, preserves the antiperiodicity in $x_0$.

Due to the fact that these eigenfunctions are eigenfunctions of $\sigma_3$, one can equivalently write gauge transformations in the form $\psi_{k,\,l}(x)\rightarrow e^{i\frac{\sigma_3}{2}2\gamma x_0}\psi_{k,\,l}(x)$. This last expression shows that, as $x_0$ grows from $0$ to $\beta$, spinors are rotated by $2\gamma\beta$, since $\frac{\sigma_3}{2}$ is the generator of rotations in the plane $x_1x_2$. In particular, $\gamma=\frac{2\pi}{\beta}$ corresponds to a $4\pi$-rotation around the magnetic field. On the other hand, $\gamma=\frac{\pi}{\beta}$ corresponds to a $2\pi$-rotation. At finite temperature, such transformation changes the statistics to a bosonic one. For $\kappa=+1$, it also gives rise to an overall phase of $\pi$ per unit degeneracy in the partition function. Such phase is the contribution which survives in the zero temperature limit. Always in the zero temperature limit, $\kappa=+1$ gives rise to a Chern-Simons term in the effective action. Invariance of the partition function under rotations of $2\pi$ requires the reduced flux ($\Delta_L$) to be an integer, which fixes the coefficient in front of the Chern-Simons term. Such term is not present for $\kappa=0$.

To summarize, in each representation, the effective action per unit degeneracy is invariant under large gauge transformations, with any of the two possible selections of phase. As a result, the invariance persists no matter which of the three possible combinations of phases is selected. Moreover, each of the two selections of phase in each representation corresponds to a different geometric phase under the rotation of spinors along a closed path around the magnetic field ($\kappa=-1$: no geometric phase; $\kappa=+1$: geometric phase of $\pi$). So, the three possible combinations of phases of the determinant then give a total phase in the partition function of $\pi$ ($K=1$, monolayer), $2\pi$ ($K=2$, bilayer), or $0$ ($K=0$), to be compared with the Berry phases studied, for instance, in \cite{kope}. Finally, we note that these three values of $K$ also correspond to the three nonequivalent unitary representations of the generator of the cyclic group $C_3$, which is the relevant symmetry in the case of free graphene.

$\phantom{aa}$\\
{\bf Acknowledgements:} I thank the organizers of SPMTP08 for a very interesting conference and for the nice atmosphere enjoyed during the event. This work was partially supported by Universidad Nacional de La Plata (Proyecto 11/X381) and CONICET (PIP 6160).

\end{document}